\documentclass[twocolumn,amsmath,amssymb,prl,footinbib,floatfix,superscriptaddress]{revtex4}
\usepackage{graphicx}
\usepackage{dcolumn}
\usepackage{bm}

\usepackage{amssymb}
\usepackage{amsmath}
\usepackage{hyperref}
\usepackage{color}
\usepackage{soul}
\usepackage{xcolor}

\definecolor{red}{rgb}{0.7,0,0}
\definecolor{green}{rgb}{0.,0.35,0.}
\definecolor{blue}{rgb}{0.2,0.2,0.7} 
\definecolor{black}{rgb}{0.15,0.15,.15}
\def\ket#1{\left| #1 \right\rangle}
\def\matrixel#1#2#3{\left\langle #1 \right| #2 \left| #3 \right\rangle}
\def\mel#1#2#3{\langle #1 | #2 | #3 \rangle}
\def\ev#1#2{\mel{#2}{#1}{#2}} 
\def\bcs{\mathrm{bcs}}
\def\scs{\mathrm{scs}}

\def\barray{\begin{eqnarray}}
\def\earray{\end{eqnarray}}
\def\beq{\begin{equation}}
\def\eeq{\end{equation}}
\def\({\left(}
\def\){\right)}

\makeindex

\begin{document}

\title{Symmetry protected phases in inhomogeneous spin chains}

\author{Nadir  Samos S\'aenz de Buruaga$^1$, Silvia N. Santalla$^2$,
 Javier  Rodr\'{\i}guez-Laguna$^{3}$ and Germ\'an  Sierra$^{1}$ \\
${}^1$ Instituto de F\'{\i}sica Te\'orica UAM/CSIC, Universidad
 Aut\'onoma de Madrid, Cantoblanco, Madrid, Spain\\
${}^2$ Dept. de F\'{\i}sica and Grupo Interdisciplinar de Sistemas
 Complejos (GISC), Universidad Carlos III de Madrid, Spain\\
${}^3$ Dept. de F\'{\i}sica Fundamental, Universidad Nacional de
 Educaci\'on a Distancia (UNED), Madrid, Spain}

\date{\today}

\begin{abstract} 
It has been shown recently that inhomogenous spin chains can exhibit
exotic phenomena such as the breaking of the area law of the
entanglement entropy. An example is given by the rainbow model where
the exchange coupling constants decrease from the center of the
chain. Here we show that by folding the chain around its center, the
long-range entanglement becomes short-range which can lead to
topological phases protected by symmetries (SPT). The phases are
trivial for bond-centered foldings, and non trivial for site-centered
ones. In the latter case, the folded spin 1/2 chain with $U(1)$
symmetry belongs to the Su-Schrieffer-Heeger class, while the folded
chain with $SU(2)$ symmetry is in the Haldane phase. Finally, we
extend these results to higher spin chains where we find a
correspondence between the symmetry protection of gapped and gapless
phases.
\end{abstract}

\maketitle

{\em Introduction.-} In the last years, an area called Quantum Matter
has emerged, where Condensed Matter Physics and Quantum Optics find a
common ground to exchange ideas and techniques. Some antecedents of
this area can be found in the 80's in the integer and fractional
Quantum Hall effects \cite{FQH} that paved the way to the more recent
discovery of topological insulators and superconductors
\cite{Bernevig2013}, Weyl semimetals \cite{Felser2015}, etc. The
description of Quantum Matter goes beyond the Landau paradigm in terms
of symmetry breaking and local order parameters. The fundamental
concept here is that of topology which in this context means that the
relevant properties of a physical system are distributed throughout
its extent, whose characterization requires the use of advanced
mathematical tools
\cite{Wen,Kennedy1992,Altland1997,GW09,P10,Fidkowski2010,Fidkowski2011,Turner2011,Schuch2011,W11a,W11b,P12,Chen2013,TW15,Bernevig2015,CS17,FO17}.

In this letter we shall focus on 1D spin systems whose topological
properties have been characterized in various ways.  Consider for
example the antiferromagnetic Heisenberg chain (AFH) of spin 1. As
famously conjectured by Haldane\cite{Haldane1983}, the AFH Hamiltonian
has a unique ground state which does not break the rotational symmetry
$SO(3)$, and has a gap in the spectrum for periodic chains. This
conjecture led Affleck et al. \cite{AKLT} to propose a state whose
properties are similar to those of Haldane's state, and whose
many-body wave function is a matrix product state (MPS)
\cite{Schollwoeck2011,Orus2014a}. The topological properties of the
Haldane and the AKLT states were characterized by an string order
parameter of den Nijs and Rommelse \cite{NR89} or a symmetry dependent
string order parameter \cite{Haegeman2012}, and the existence of
effective spins 1/2 at the ends of an open chain. It was realized that
the Haldane phase can be protected by several symmetries like $Z_2
\times Z_2$ \cite{Kennedy1992}, time reversal and inversion symmetry
\cite{GW09,P12}, which guarantee \emph{independently} the degeneracy
of the entanglement spectrum. More importantly, the concept of
symmetry protection turns out to be the key to understand and classify
the phases with short range entanglement where one can apply the MPS
techniques \cite{Schuch2011,W11b}.

Here we present a new way to generate symmetry protected phases in 1D
using local Hamiltonians that are inhomogeneous and without a gap in
the spectrum. At first glance, one does not expect this possibility to
occur because the corresponding ground states would develop long-range
entanglement that violates the area-law
\cite{Hastings2006,Wolf2008,VR10}. However, we will show that a
rearrangement of the sites transforms the long-range entanglement into
a short-range entanglement, where standard methods can be applied to
identify the possible phases \cite{W11b,Chen2013}. The inhomogeneous
models that we will consider are obtained by a deformation of the
critical models where the entropy of entanglement scale
logarithmically \cite{LT,CC}. The effect of the lack of homogeneity is
to increase this violation that becomes linear in the size of the
blocks, like a thermal entropy. This mechanism has a geometrical
interpretation in the underlying conformal field, according to which
inhomogeneity corresponds to a curvature of space-time
\cite{RD17,TR18}.

A byproduct of our construction is that it suggests a relationship
between the SPT phases and the phases described by conformal field
theories (CFT) in terms of global anomalies \cite{FO17}. The reason is
that the former are constructed from a special deformation of the
latter. We shall illustrate this relation with several examples.

\begin{figure}
 \includegraphics[width=8.5cm]{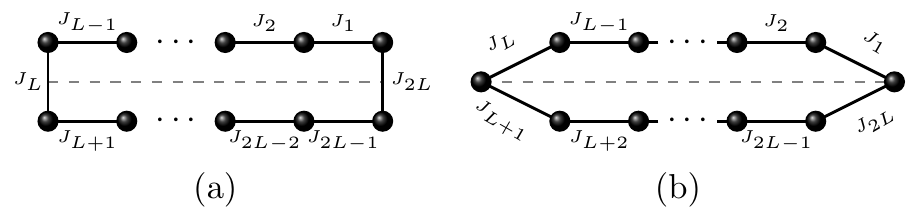}
 \caption{Illustrating our physical model, Eq. \eqref{eq:ham} with
  couplings given in Eq. \eqref{eq:Jvalues}. Symmetrical links with
  respect to the dashed line carry the same couplings. (a)
  Bond-centered symmetry (bcs); (b) Site-centered symmetry (scs). }
 \label{fig:illust}
\end{figure}

{\it The rainbow XX model.-} We start with the inhomogeneous XX spin
chain whose exchange coupling constants decrease exponentially from
the center \cite{VR10,RR14,RR15,RD17,TR18,AS18}. This model is
equivalent to a spinless fermion model by a Jordan-Wigner
transformation. The Hamiltonian for a closed chain with an even number
of sites, $2L$, is
\barray 
H & = & - \frac{1}{2} \sum_{n=1}^{2 L} J_n c_{n}^\dagger c_{n+1} + h.c.  \; 
\label{eq:ham} 
\earray 
where $c_n$ is the fermion annihilation operator, and $c_{2L+1} =
c_1$. We define models with bond-centered symmetry (bcs) and
site-centered symmetry (scs) as those satisfying
\barray 
J_n =  J_{2L-n} \; ({\rm bcs}), \quad 
J_n =  J_{2L+1-n} \; ({\rm scs}) \label{2} 
\earray 
that correspond to inversions around bond $(L, L+1)$ and site $L+1$
respectively. The lack of homogeneity is introduced in terms of an
exponential decrease of the hopping parameters from the center of the
chain outwards,
\barray 
J_{n \neq L} &= & e^{ - h |n - L|}, \;
J_L = e^{ - \frac{h}{2}} \; \;  ({\rm bcs}), \label{eq:Jvalues} \\
J_{n \neq L, L+1} & = & e^{ - h (\left|n - \(L + \frac{1}{2}\)\right| - \frac{1}{2})}, \;
J_L = J_{L+1} = e^{ - \frac{h}{2}} \; \;  ( {\rm scs}) \, ,  \nonumber 
\earray 
where $h \geq 0$ is the inhomogeneity parameter. The symmetries of the
two types of chains are illustrated in Fig. \ref{fig:illust}. In the
bcs model the highest coupling lies at the center of the chain, that
is $J_L$, while the weakest coupling is $J_{2L} = e^{-hL}$ that
connects sites 1 and $2L$. In the strong inhomogeneity limit $hL \gg
1$, one can set $J_{2L}$ to zero which leads to the rainbow chain
studied in references \cite{RR14,RR15,RD17,TR18,AS18}. Using the
Dasgupta-Ma RG method \cite{DM,F} it was shown that the ground state
of the rainbow chain takes the form \cite{RR14}
\beq
|{\rm bcs}\rangle
\stackrel{h \rightarrow \infty}{\longrightarrow} 
d_{+1}^\dagger d_{-2}^\dagger d_{+3}^\dagger \dots
d_{\eta_L L}^\dagger | 0 \rangle \, , 
\label{4}
\eeq 
where
\beq
d_{\pm n} = \frac{1}{ \sqrt{2}} ( c_{L +1 - n} \pm c_{L+ n}), \quad
n=1, \dots, L \, ,
\label{5}
\eeq
are fermion operators on the opposite sides of the chain, that
annihilate the Fock vacuum $|0 \rangle$ and $\eta_L = (-1)^{L+1}$.
This state presents a maximal violation of the entanglement entropy
for the block $A = \{1, 2, \dots, \ell \}$:
\beq
S^\bcs_A = \ell \ln 2 , \qquad \ell \leq L \, . 
\label{6}
\eeq
Let us consider now chains with site-centered symmetry. In the limit
$h \gg 1$, the dominant interaction takes place between sites $L, L+1$
and $L+2$. In this situation one should use first order perturbation
theory to renormalize three spins into one effective spin
(Supplementary Material). Iterating this RG procedure one obtains the
ground state,
\beq
|{\rm scs} \rangle
\stackrel{h \rightarrow \infty}{\longrightarrow}
b_{+0}^\dagger b_{-1}^\dagger b_{+2}^\dagger b_{-3}^\dagger \dots
b_{+  L}^\dagger | 0 \rangle \, , 
\label{7} 
\eeq\\
where 
\barray 
f_0 & = & c_{L+1} , \quad f_L = c_{1} \label{8} \\ 
f_{\pm n} & = &  \frac{1}{ \sqrt{2}} ( c_{L +1 - n} \pm c_{L+ 1+ n}), \quad n=1, \dots, L -1 \, , \nonumber \\ 
b_{\pm 0} &  = &  \frac{1}{ \sqrt{2}} ( f_0  \pm f_{+ 1} ), \quad 
b_{\pm L}  =   \frac{1}{ \sqrt{2}} ( f_0  \pm f_{+ L-1} ), \nonumber \\
b_{\pm n} & =  &  \frac{1}{ \sqrt{2}} ( f_{\pm n} + f_{\pm (n+1)} ) \, . \nonumber 
\earray 
The entanglement entropy of the block $A$ is (SM) 
\beq
S^\scs_A =
\left( \ell + 1 \right) (2  \ln 2 - 1) , \qquad \ell \leq L \, , 
\label{9}
\eeq
which is still linear but not maximal as in Eq.\eqref{6}.

\begin{figure}[h!]
\includegraphics[width=85mm]{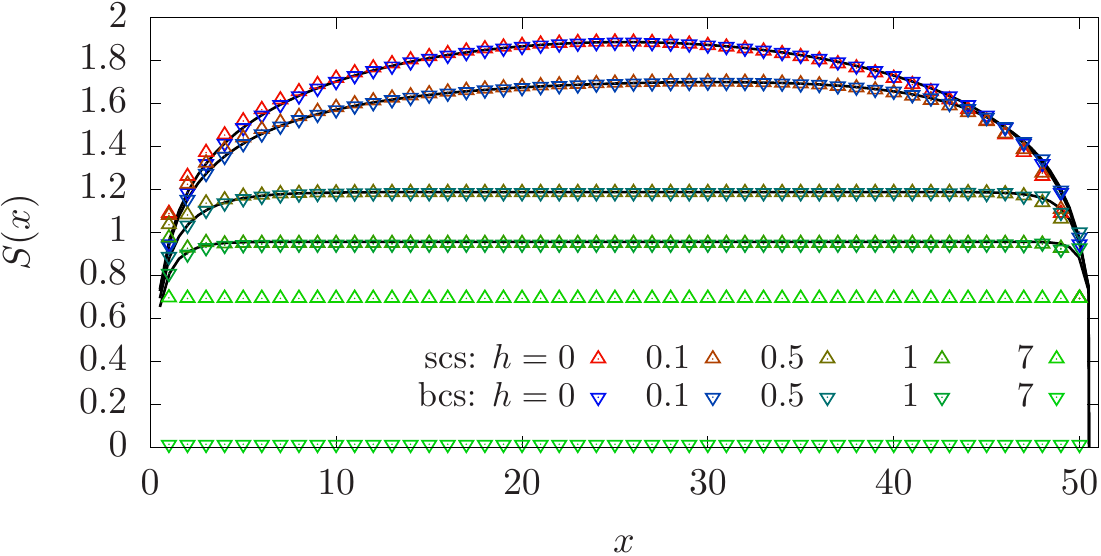}
\caption{Entanglement entropy of the central block for increasing
  values of $h$ shown in descending order, of the bcs and scs chains
  with $L=51$. The continuum lines are the CFT prediction \eqref{13}.}
\label{central}
\end{figure}

The long-range entanglement of the bcs/scs states can be converted
into short-range one using the basis of states generated by the
operators $d_{\pm, n}$ and $f_{\pm, n}$. The $d$-operators are the
bonding and anti-bonding combinations of the fermions located at
opposite sites in the chain. They become local operators by {\em
  folding} the chain around the bond $(L, L+1)$ that transforms it
into a ladder with 2 legs and $L$ rungs. The folding {\em trick} has
played an important role in the study of quantum impurity problems
\cite{KF92,SA01,BF12}. Equation \eqref{4} shows that the bcs state is
the product of bonding and antibonding states on the rungs. Hence, the
entanglement entropy of the block $C = \{ L+1- \ell, \dots, L+ \ell
\}$ located at the center of the chain, that corresponds to $\ell$
rungs in the ladder, is simply
\beq
{S}^\bcs_{C} = 0, \quad  {\ell}=1, \dots, L-1 \, . 
\label{10}
\eeq 
Turning to the site-centered chains we observe that the $f$-operators,
Eq. \eqref{8}, involve a folding transformation that leaves sites 1
and $L+1$ untouched. The chain is now transformed into a ladder with
$L-1$ rungs and two isolated sites on both edges. The entanglement
entropy of block $C = \{ L+ 2- \ell, \dots, L+ \ell \}$ located at the
center of the chain (see Fig. \ref{fig:illust}), that corresponds to
$\ell$ rungs in the ladder and one site, is given by
\beq
{S}^\scs_{C} = \ln 2, \quad  {\ell}=1, \dots, L-1 \, . 
\label{11}
\eeq 
\begin{figure}
\includegraphics[width=70mm]{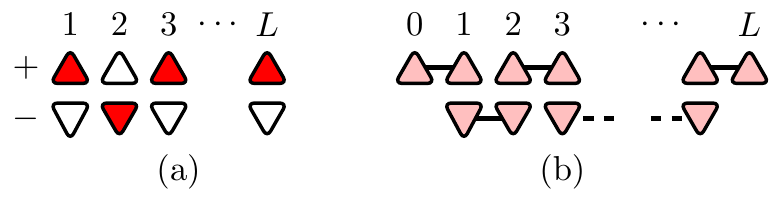}
\caption{Ground states of the rainbow XX model in the limit $h
  \rightarrow \infty$. We use the folded representation of
  Fig. \ref{fig:illust}.}
\label{even-odd}
\end{figure}
One can verify Eqs.(\ref{10}) and (\ref{11})  by looking at Fig. \ref{even-odd} which
shows that the GS of the bcs rainbow is a charge density wave (CDW),
while that for the scs rainbow is a staggered dimer state, reminiscent
of the trivial and topological phases of the well known SSH model
\cite{SSH}. The origin of these GS structures can be understood
writing the Hamiltonian Eq. \eqref{eq:ham} using the operators
\eqref{5} and \eqref{8},
\def\ds{\displaystyle}
\barray 
&H_\bcs = \ds - \frac{1}{2} \left[\sum_{n=1}^{L-1} J_{L-n} (
 d^\dagger_{+n} d_{+n+1} + d^\dagger_{-n} d_{-n+1}) \right.
 \nonumber \\ 
 &+ J_L ( d^\dagger_{+1} d_{+1} - d^\dagger_{-1} d_{-1} ) \nonumber \\
 &+ \left. J_{2L} ( d^\dagger_{+L} d_{+L} - d^\dagger_{-L} d_{-L} )
 + {\rm h.c.} \right],   \nonumber \\
&H_\scs = \ds - \frac{1}{2} \left[
 \sum_{n=1}^{L-2} J_{L-n} ( f^\dagger_{+n} f_{+n+1} +
 f^\dagger_{-n} f_{-n+1} ) \right. \nonumber \\
 &\left. +\sqrt{2}\, J_L\; f^\dagger_0 f_{+1}
 + \sqrt{2}\, J_1\; f^\dagger_L f_{+ L-1} + {\rm h.c.}\right]  \, . 
\label{eq:Hbcsscs}
\earray 
In the bcs model the chemical potential on the first rung induces, in
the strong inhomogeneity limit, the full occupation of site $+1$ and
the emptying of site $-1$. This mechanism gives rise to a CDW
state. In the scs model, the hopping term between the isolated mode
$f_0$ and the mode $f_{+1}$ of the first rung, induces in the same
limit, a hybridization that propagates along the chain producing a
staggered dimer state on the ladder. These ground states are
illustrated in Fig. \ref{even-odd}. The bcs state is a product state
in the basis $d_{\pm, n}$, so a MPS with bond dimension $\chi=1$. On
the other hand, the scs state is a product of dimers, with bond
dimension $\chi=2$ (SM). The entanglement spectrum is twofold
degenerate with two equal eigenvalues. We shall next show that the
previous topological features persist for all values of $h
>0$. Fig. \ref{central} shows the entanglement entropies (EE) of the
central blocks $A= [-x, x]$ for the bcs and scs rainbows. They are
independent of $x$, for sufficiently large values, so corresponding to
an area law for the folded chain. Notice that when $h \gg 1$ the EE
for the bcs chains goes to zero while that for the scs chains goes to
$\ln 2$, in agreement with Eqs. \eqref{10} and \eqref{11}. The rainbow
chain has a continuum limit given by a massless Dirac fermion on a
curved spacetime. Using CFT techniques one can find the entanglement
entropy of the central block \cite{RD17} $A =[-x,x]$ (with   $x=n- L- \frac{1}{2}$ 
for bcs chains and $x = n - L -1$ for the scs chain)
is
\beq
{S}_{C}(x) \simeq \frac{1}{3} \ln \left[ \frac{ 4 \tilde{L}}{\pi} e^{- h x} \sin \frac{ \pi \tilde{x}}{\tilde{L}} \right] + E_1  < 
\frac{1}{3} \ln \frac{4}{h} + E_1
\label{13}
\eeq
where $\tilde{x} = ( e^{h x}-1)/h$ and $E_1 \simeq 0.49502$.
Fig.\ref{central} shows that this expression reproduces the bcs and
scs entropies when $h$ is not too large, where the field theory limit
applies.  The upper bound in Eq.\eqref{13} is similar to the EE of a
massive theory in the scaling limit with $1/h$ as correlation length
\cite{CC}.

\begin{figure}
\includegraphics[width=60mm,angle=0]{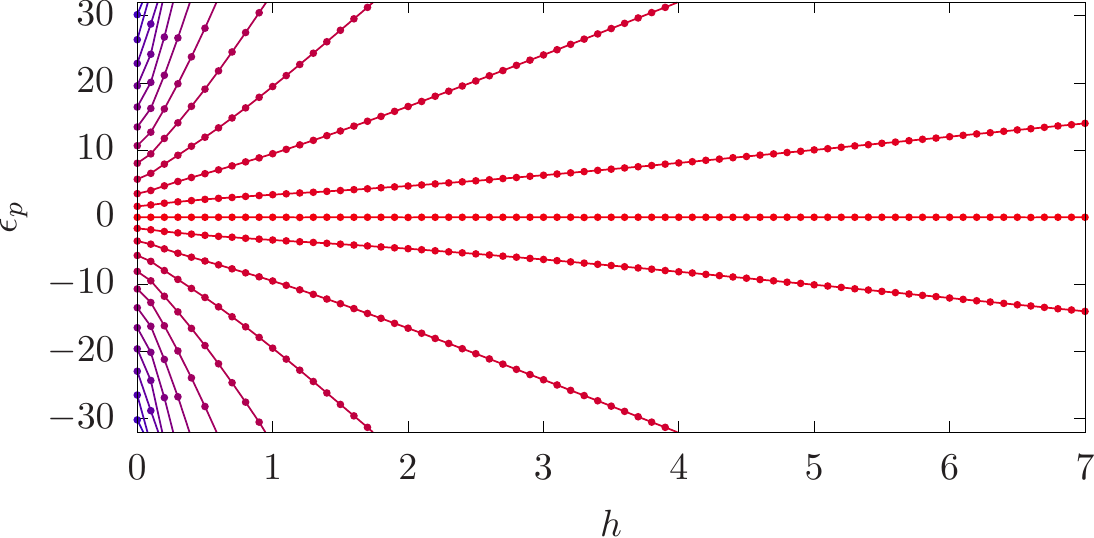}
\caption{Single particle entanglement spectrum $\varepsilon_p$ for a
 scs chain with $L=51$ as a function of $h$. }
\label{ES-good}
\end{figure}

Another signature of a SPT phase is the degeneracy of the entanglement
spectrum (ES) \cite{P10,Turner2011}. For a free fermion system the
entanglement energies are given by $E(\{ n_p \}) = \sum_p
\varepsilon_p n_p + r_0$, where $\{ n_p =0, 1 \}$ is the set of
occupation numbers of the one-body entanglement entropies
$\varepsilon_p$ that are computed from the eigenvalues of the
correlation matrix $\langle c^\dagger_n c_m \rangle$
\cite{Peschel2003}. For scs chains there exists a zero mode,
$\varepsilon_0 =0$, for all values of $h$, that gives rise to a doubly
degeneracy of the ES (see Fig. \ref{ES-good}). This degeneracy is
protected by the time reversal and particle-hole symmetry of the
Hamiltonian. Hence, this model belongs to the symmetry class BDI of
topological invariants for free fermions, the same as the SSH model
\cite{SSH,Fidkowski2011,Bernevig2015}: a perturbation to the
Hamiltonian that does not respect those symmetries will break the ES
degeneracy.

\begin{figure}
\includegraphics[width=70mm, angle = 0]{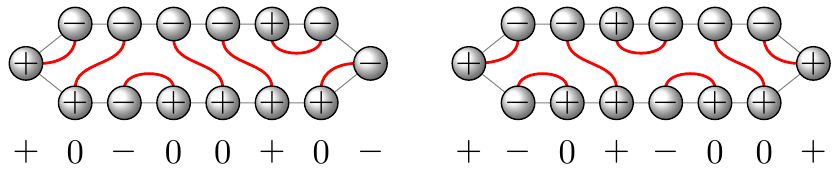} 
\caption{Strong inhomogeneity limit of the GS of the scs Heisenberg
  model.  The links represent valence bonds. The $\pm$ symbols inside
  the balls denote the sign of $\sigma^z$ of the corresponding spin,
  while the sign of the sum over the rung appears below. These signs
  display an antiferromagnetic liquid behaviour characteristic of the
  Haldane phase. }
\label{fig:leopard}
\end{figure}

\begin{figure}
\includegraphics[width=80mm, angle = 0]{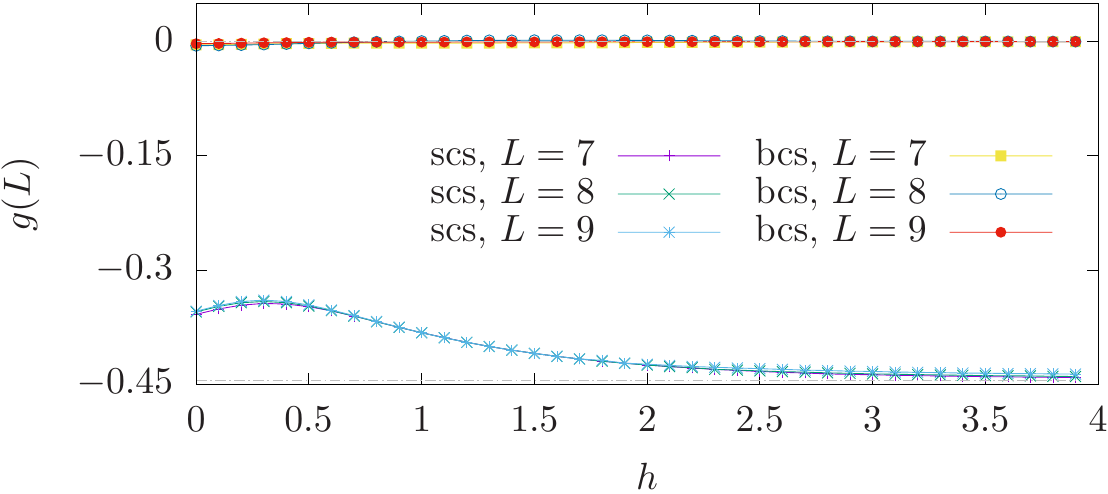}
\\ \includegraphics[width=80mm, angle = 0]{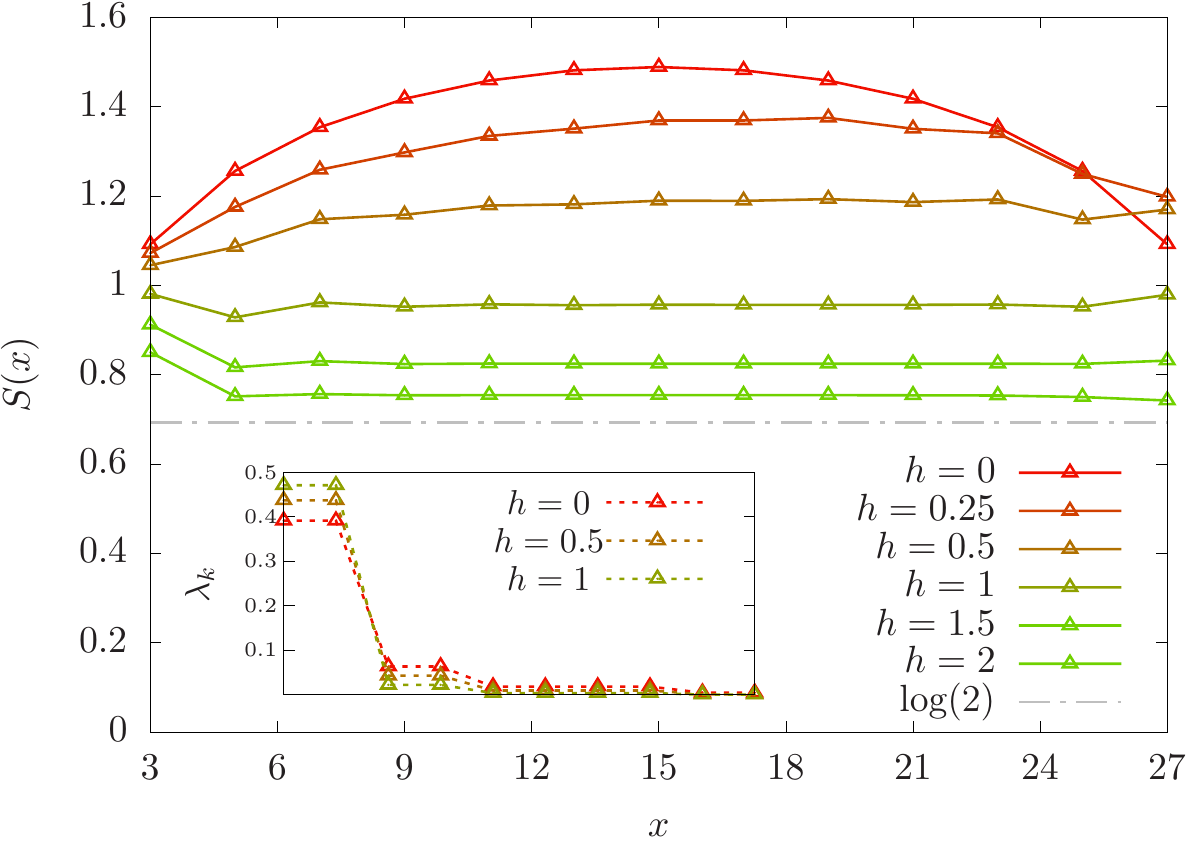}
\caption{ Top: Plot of the string order parameter, $g(L)$, for the
  Heisenberg chains with $L=7,8,9$. It vanishes for the bcs chains and
  approaches the AKLT value $-4/9$ for the scs chains. Bottom: EE of
  the central blocks of the scs Heisenberg model with $L=15$ obtained
  with the DMRG \cite{White1992} as a function of the size and several
  values of $h$. Notice the convergence to $\log 2$ already for
  $h=2$. Inset: ES as a function of the order for 3 values of $h$ and
  central block with $x =15$ sites.}
\label{sop}
\end{figure}

{\em The rainbow antiferromagnetic Heisenberg model.-} The Hamiltonian is 
\barray 
H & = & \sum_{n=1}^{2L} J_n\; {\bf S}_{n} \cdot {\bf S}_{n+1} \, , 
\label{eq:heis} 
\earray 
where ${\bf S}_{n}$ are the spin 1/2 matrices at site $n$. The
couplings $J_n$ are defined in Eq. \eqref{eq:Jvalues}. Let us study
the phases of this model in the limit $h \gg 1$. For the bcs chain,
the analysis is similar to the one of the bcs XX chain. The
Dasgupta-Ma RG equation yields a GS made of spin singlets between
sites $n$ and $2L+1-n$. Folding the chain maps this state into the
product of $L$ rung singlets of the two leg ladder. In the scs chain,
the highest couplings are $J_L = J_{L+1}$, and we start diagonalizing
the Hamiltonian $J_L {\bf S}_{L+1} \cdot ( {\bf S}_{L} + {\bf S}_{L+2}
)$. Its GS is obtained by forming a triplet between spins ${\bf
  S}_{L}$ and ${\bf S}_{L+2}$, that couples to spin ${\bf S}_{L+1}$
yielding an effective spin 1/2, denoted as ${\bf S}'_{L+1}$. First
order perturbation theory yields the RG equations ${\bf S}_{L}$, ${\bf
  S}_{L+2} \rightarrow \frac{2}{3} {\bf S}'_{L+1}$. The next order
term in the Hamiltonian is $J_{L-1} ( {\bf S}_{L-1} \cdot {\bf S}_{L}
+ {\bf S}_{L+2} \cdot {\bf S}_{L+3} )$ that gets renormalized into
$\frac{2}{3} J_{L-1} {\bf S}'_{L+1} \cdot ( {\bf S}_{L-1} + {\bf
  S}_{L+3} )$, so we can repeat the same RG step done above if $h \gg
1$. Each RG step generates an effective spin 1 that couples to an
effective spin 1/2 from the previous step. Completing the RG procedure
yields a chain with $L$ effective spins 1 and two spins 1/2 at the
ends of the folded chain (see Fig. \ref{fig:leopard}). This is the
AKLT \cite{AKLT} state of an open chain with $L-1$ spins 1's and two
1/2's at the ends (SM).
The RG method used above is valid for $h \gg 1$, but the topological
nature of the GS also holds for all positive values of $h$. To verify
this statement we show in Fig. \ref{sop} the string order parameter
\cite{NR89,GA89,WH93}.
\beq
g(L) = \langle S^z_{1} \, e^{ i \pi \sum_{j=2}^{L-1}  S^z_j} \, S^z_{L} \rangle, 
\label{eq:sop}
\eeq
where $S^z_j= S^z_{u,j} + S^z_{d,j}$ is the spin operator on the
$j^{\rm th}$- rung of the folded chain. For bcs chains all rungs are
considered, while for scs chains sites $1$ and $L+1$ are left out. For
the bcs chains, $|g(L)|$ approaches quickly zero as $h \gg 1$, while
for the scs chains $g(L)$ converges asymptotically towards $-4/9$ that
corresponds to the AKLT state \cite{NR89}. Fig. \ref{sop}-bottom,
shows that the EE of the central blocks of the scs model remains
constant for sufficiently large values of $h$, which is a signature of
the area law (recall Eq.\eqref{13}). The entanglement spectrum is
doubly degenerate (see inset of Fig. \ref{sop} bottom), that is
another feature of the SPT phase, which in this case is protected by
the time reversal symmetry \cite{P10,P12}. Moreover, if we drop the site $2N$,
that is placed in the right most position in Fig. \ref{fig:leopard}, this has the effect of
leaving an edge spin of the effective spin 1 chain.

{\em Gapless versus gapped topological phases.-} The previous results
show that a strong inhomogeneous deformation of the critical spin 1/2
AFH chain, with site centered symmetry, becomes an effective spin 1
chain in the Haldane phase. Let us now consider Heisenberg chains with
higher spin. If the spin is a half-odd integer, $S= \frac{1}{2},
\frac{3}{2}, \dots$, then the uniform AFH Hamiltonian is gapless and
described by the $SU(2)_1$ WZW model \cite{Affleck1987a}. Applying a
strong inhomogeneous scs deformation generates, via its folding, an
effective AKLT chain with spin $2S=1,3, \dots$. The ground states of
these spin chains possess non-trivial SPT phases \cite{P12}. Let us
now replace the AFH Hamiltonian by the Babujian-Takthajan (BT)
Hamiltonian \cite{Takhtajan1982,Babudjian1982,Babudjian1983} of spin
$S$ that is integrable and described by the $SU(2)_k$ WZW model with
$k = 2S$. We also expect its strong scs deformation to map into the
AKLT state of spin $2S$. Hence, when $k$ is odd, the AFH and BT models
both end up in non-trivial SPT phases. Repeating this process for
integer spin chains gives trivial SPT phases. Indeed, the AFH
Hamiltonian for integer spin is gapped according to Haldane's
conjecture \cite{Haldane1983}. Its strong scs deformation gives an
AKLT state with spin $2S=2,4, \dots$ which is a trivial SPT phase
integer \cite{P12}. The same is expected to hold for the scs
deformation of the BT model for integer spin. The difference between
even and odd levels of critical spin chains with $SU(2)_k$ symmetry
reminds the one based on global anomalies that also lies on the parity
of $k$ \cite{FO17}. The mechanism of relating apparently different
phases via inhomogeneities can be extended to other model in 1 and 2 dimensions.
 An example is the spin 1/2 AFH model on a square lattice. A
strong site-centered deformation of the exchange couplings along the
$X$ and $Y$-axes yields a two dimensional AKLT state with spins 2 in
the bulk, spins 1 along the edges and spins 1/2 at the corners. 

We want also to mention the proposed relation between SPT phases, boundary CFT
and entanglement entropies that was proposed recently in reference \cite{Ryu}. 

In summary, we have found  a new mechanism to generate symmetry
protected phases in one dimensional spin chains governed by  inhomogeneous local
Hamiltonians.  The ground states of these models  
have long range entanglement but folding the chains around their center
its becomes short range. 
We illustrate this method with the spin 1/2 XX and antiferromagnetic 
Heisenberg chains whose inhomogeneous deformations, with site-centered symmetry,
yields ground states in the  SSH and Haldane phases respectively.   
We expect this mechanism to work for other 1D and 2D models, that 
poses the question of which SPT and topological phases can be constructed 
playing this {\em origami}  game, and whether they could be realized
experimentally for example by  applying pressure to real  materials, or  in synthetic materials realized
in optical lattices  \cite{optical}.

\vspace{2mm}
\hrule
\vspace{2mm}

{\em Acknowledgements.} 
\begin{acknowledgments}
We would like to thank V. Alba, I. Bloch, P. Calabrese, X. Chen,
J. I. Cirac, A. Feiguin, E. Kim, J. I. Latorre, E. L\'opez, F. Mila,
D. P\'erez Garc\'{\i}a, G. Ram\'{\i}rez, S. Ryu, E. Tonni and
H. Yarloo for conversations. We acknowledge financial support from the
grants FIS2015-69167-C2-1-P, FIS2015-66020-C2-1-P, QUITEMAD+
S2013/ICE-2801 and SEV-2016-0597 of the ``Centro de Excelencia Severo
Ochoa'' Programme.
\end{acknowledgments}



\newpage
\onecolumngrid

\section*{\large\bf{SUPPLEMENTAL MATERIALS}}
\section{XX model} 
\label{sec:xx_model}

\subsection{Single-Body Modes}

While the bcs chain (see Eq. \eqref{eq:ham} and
Eq. \eqref{eq:Jvalues}) is tractable via the strong-disorder
renormalization group (SDRG) \cite{DM,F}, the method is inconclusive
for the scs chain. The reason is that there are two hoppings of equal
magnitude $\exp(-\frac{h}{2})$. Even if we choose one of them randomly
to put a valence bond on it, the degeneracy will propagate to the next
renormalization step. Thus, we have developed a different approach via
a real space renormalization method \`a la Wilson, based on the
single-particle character of this problem.

Numerical studies of the GS of the scs rainbow system show that the
single-body modes are localized in the strong-inhomogeneity limit, but
on {\em four} sites. Moreover, as the modes increase in energy, their
support moves outwards from the center of the chain. This leads to a
natural renormalization scheme which starts out with the central
block, $B^{(1)}$, comprising the three central sites: $\bullet_{L-1}$,
$\bullet_L$ and $\bullet_{L+1}$. The two internal couplings are the
same, equal to $\exp(-h/2)$, so the effective Hamiltonian is:

\begin{equation}
H^{(1)}=-e^{-h/2}\left(\begin{matrix}
0 & 1 & 0 \\
1 & 0 & 1 \\
0 & 1 & 0 \\
\end{matrix}\right).
\end{equation}
Its spectrum is composed of three values, $E_i$, with their associated
eigenvectors, $\ket{i^{(1)}}$ where $i\in \{-,0,+\}$. Let us select the
ground state, $E_-$, which has the form:

\begin{equation}
\ket{-^{(1)}}=\frac{1}{2}(1,\sqrt{2},1),
\end{equation}
and keep it as the first electronic orbital. Then we proceed to take
the zero mode,

\begin{equation}
\ket{0^{(1)}} = \frac{1}{\sqrt{2}}(1,0,-1),
\end{equation}
and take it to the next RG level, along with the orbitals located on
the neighboring sites to the block: sites $\bullet_{L-2}$ and
$\bullet_{L+2}$. These three single-body orbitals:
$\ket{\bullet_{L-2}}$, $\ket{0^{(1)}}$ and $\ket{\bullet_{L+2}}$
constitute block $B^{(2)}$. Let us build the effective Hamiltonian:

\begin{equation}
H^{(2)}=\left(\begin{matrix}
\matrixel{\bullet_{L-2}}{H}{\bullet_{L-2}} &
\matrixel{\bullet_{L-2}}{H}{0^{(1)}} &
\matrixel{\bullet_{L-2}}{H}{\bullet_{L+2}} \\
\matrixel{0^{(1)}}{H}{\bullet_{L-2}} &
\matrixel{0^{(1)}}{H}{0^{(1)}} &
\matrixel{0^{(1)}}{H}{\bullet_{L+2}} \\
\matrixel{\bullet_{L+2}}{H}{\bullet_{L-2}} &
\matrixel{\bullet_{L+2}}{H}{0^{(1)}} &
\matrixel{\bullet_{L+2}}{H}{\bullet_{L+2}}
\end{matrix}\right)
\end{equation}
The lowest energy eigenstate $\ket{-^{(2)}}$ is kept as a new orbital,
and there appears a new zero mode, $\ket{0^{(2)}}=
\frac{1}{\sqrt{2}}(1,0,1)$, which is taken to the next RG level. The
$n$-th RG step is predicated on a block $B^{(n)}$ which contains the
zero mode of the previous step, $\ket{0^{(n-1)}}$ and the next two
site-orbitals, $\ket{\bullet_{L-n}}$ and $\ket{\bullet_{L+n}}$, with
effective Hamiltonian:

\begin{equation}
H^{(n)}=\left(\begin{matrix}
\matrixel{\bullet_{L-n}}{H}{\bullet_{L-n}} &
\matrixel{\bullet_{L-n}}{H}{0^{(n-1)}} &
\matrixel{\bullet_{L-n}}{H}{\bullet_{L+n}} \\
\matrixel{0^{(n-1)}}{H}{\bullet_{L-n}} &
\matrixel{0^{(n-1)}}{H}{0^{(n-1)}} &
\matrixel{0^{(n-1)}}{H}{\bullet_{L+n}} \\
\matrixel{\bullet_{L+n}}{H}{\bullet_{L-n}} &
\matrixel{\bullet_{L+n}}{H}{0^{(n-1)}} &
\matrixel{\bullet_{L+n}}{H}{\bullet_{L+n}}
\end{matrix}\right) \quad n=1,...,L-1
\end{equation}
%
The last step of the procedure is different: the new block is built up
with the zero mode of the previous step, but there is only one
remaining orbital. Hence, the effective block is a $2\times 2$ matrix:

\begin{equation}
H^{(L)}=\left(\begin{matrix}
\matrixel{0^{(L-1)}}{H}{0^{(L-1)}}&
\matrixel{\bullet_{2L}}{H}{0^{(L-1)}} \\
\matrixel{0^{(L-1)}}{H}{\bullet_{2L}} &
\matrixel{\bullet_{2l}}{H}{\bullet_{2L}}
\end{matrix}\right).
\end{equation}
$H^{(L)}$ has two different forms depending of nature of last zero
mode: if $N\equiv0\pmod4$, $\ket{0^{L-1}}$ is symmetric while if
$N\equiv2\pmod4$ it is antisymmetric. Hence, the energy spectrum
presents a double degeneracy of $E=0$ in th former case, while in the
latter it does not.

This RG procedure allows to obtain corrections in $h$ on the step $n$
by choosing the eigenvalue $E^{(n)}_-$. As a consequence of the growth
from the center along the RG process, the method can provide
corrections to the energy of every single-body mode in subsequent RG
steps. Hence, the single-body operator $b_{1}$ that appears in the
first RG step receives corrections at every RG step. Let us present
the single-body modes computed in first order in $\alpha$. Due to the
periodic boundary conditions, the mode $b^\dagger_L$ is, without
corrections, the same as $b_1^\dagger$.

\begin{eqnarray}
\footnotesize
\label{eq:orbitalop}
b_1^\dagger&=&
\left(\frac{1}{\sqrt{2}}-\frac{e^{-h}}{8\sqrt{2}}\right)c^\dagger_L+
\left(\frac{1}{2}-\frac{e^{-h}}{16}\right)
\left(c^\dagger_{L-1}+c^\dagger_{L+1}\right)
+\frac{e^{-\frac{h}{2}}}{\sqrt{2}}
\left(\sum_{i=1}^{L-2}\left(\frac{e^{h(2-L+i)}}{\sqrt{2}}\right)^{L-i}
\left(c_i^{\dagger}+c_{2L-i}^{\dagger}\right)+
2\left(\frac{e^{\frac{L-2}{2}}}{\sqrt{2}}\right)^Lc^\dagger_{2L}\right),
\nonumber\\ 
b_k^\dagger&=&
\frac{1}{2}\left(c_{L+1-k}^\dagger+c_{L-k}^\dagger+
(-1)^{k+1}(c_{L-1+k}^\dagger+c_{L+k}^\dagger)+
\sum_{l=1}^{L-2}e^{-\frac{h}{2}l(l+1)}
\left(c_{L-k-i}^\dagger+(-1)^{k+1}c_{L+k+i}^\dagger \right)\right),
\quad k=2,\cdots,L-1, \nonumber\\
b_L^\dagger&=&\frac{1}{2}\left(c_{1}^\dagger-c_{2L-1}^\dagger\right)+
\frac{1}{\sqrt{2}}c^\dagger_{2L},
\end{eqnarray}
%
where $c^\dagger_i$ is the fermionic creation operator on site
$i$. Notice that, due to particle-hole symmetry, each of these modes
has a negative energy counterpart. We would like to stress that the
physics of the strong coupling can be very well understood by
considering zero order on $\alpha$.

\subsection{Entanglement Entropy}
\label{sub:entanglement_entropy}

In this section we derive expression \ref{9} of the entanglement
entropy of a block $B$ with $l$ sites of a scs chain. It is well known
\cite{Peschel2003} that the spectrum of the reduced density matrix of
fermionic and bosonic lattice systems can be obtained through the
diagonalization of the block correlation matrices (CM), $C_{ij}$:

\begin{equation}
 C_{ij}= \ev{c^\dagger_ic_j}{GS}
 = \sum_{k,k'\in\Omega_{GS}}U_{k,i}\bar{U}_{k',j}\ev{b^\dagger_kb_{k'}}{GS}
 =\sum_{k\in\Omega_{GS}}U_{k,i}\bar{U}_{k,j},
\label{eq:corr}
\end{equation}
where $\ket{GS}$ is the ground state, $U_{k,i}$ is the unitary matrix
that diagonalizes the single-body Hamiltonian and $\Omega_{GS}$ is the
set of the negative single body levels that are ocuppied. Note that
the ground state is degenerate for scs chains with $N\equiv0\pmod4$,
since there is a (double) zero mode in the single-body spectrum. Hence
the half-filling is not well defined. In the rest of this section and
on the main text we have restricted ourselves to chains with no
degeneracy.

\begin{figure}
\begin{eqnarray}
C=\frac{1}{4}\left(
\begin{array}{cccccccccccc}
 2 & 1 & 0 & 0 & 0 & 0 & 0 & -1 & 0 & \sqrt{2}\\
 1 & 2 & 1 & 0 & 0 & 0 & 1 & 0 & -1 & 0 \\
 0 & 1 & 2 & 1 & 0 & -1 & 0 & 1 & 0 & 0\\
 0 & 0 & 0 & 2 & \sqrt{2} & 0 & -1 & 0 & 0 & 0\\
 0 & 0 & 0 & \sqrt{2} & 2 & \sqrt{2} & 0 & 0 & 0 & 0 \\
 0 & 0 & -1 & 0 & \sqrt{2}& 2 & 1 & 0 & 0 & 0 \\
 0 & 1 & 0 & -1 & 0 & 1 & 2 & 1 & 0 & 0\\
 -1 & 0 & 1 & 0 & 0 & 0 & 1 & 2 & 1 & 0\\
 0 & -1 & -1 & 0 & 0 & 0 & 0 & 1 & 2 & \sqrt{2}\\
 \sqrt{2} & 0 & 0 & 0 & 0 & 0 & 0 & 0 & \sqrt{2} & 2
\end{array}
\right)
\end{eqnarray}
\caption{Correlation matrix of a $L=5$ scs chain in the strong
  coupling limit ($h\rightarrow\infty$) computed via
  Eq. \eqref{eq:corr} using the single body modes
  Eq. \eqref{eq:orbitalop}.  }
\label{fig:corr_matrix}
\end{figure}

The computation of the entropies of the block $B$ requires the
diagonalization of the corresponding $l\times l$ block of the CM. In
the $h\rightarrow\infty$ limit, when the block $B$ does not include
any of the sites $2L,L$ or $L\pm1$, the submatrix of the CM is
tridiagonal and translational invariant (see
Fig. \ref{fig:corr_matrix}). The eigenvectors have the following form:

\begin{equation}
\label{eigen}
\ket{\varphi_k}=\sum_{m=1}^{l}\phi_m=\sum_{m=1}^{l}Ae^{ikm}+Be^{-ikm}.
\end{equation}
The eigenvalues are given by:

\begin{equation}
\lambda_k=\frac{1}{2}\left(1+\cos{k}\right).
\end{equation}
It is straightforward to obtain the dispersion relation imposing the
boundary conditions:

\begin{eqnarray}
\label{eq:knf}
\sin(k(l+1))=2m\pi \rightarrow k=\frac{2m\pi}{N+1}.
\end{eqnarray}
The von Neumann entropy is given by:

\begin{equation}
S_{odd}(l)=-\sum_k\lambda_k\log{\lambda_k}+\left(1-\lambda_k\right)\log{\left(1-\lambda_k\right)}
\end{equation}
This finite sum can be evaluated using the Euler-McLaurin formula,
inserting a finite width in momentum space, $\Delta_k=
\frac{\pi}{l+1}$. We find:

\begin{equation}
 S_\scs(l) = -\frac{l+1}{\pi}\int_0^\pi
 \(\cos^2\frac{k}{2}\log\cos^2\frac{k}{2} + \sin
 ^2\frac{k}{2}\log\sin ^2\frac{k}{2} \) dk =
 (l+1)(2\log2-1).
\end{equation}

\begin{figure}
\includegraphics[width=80mm, angle = 0]{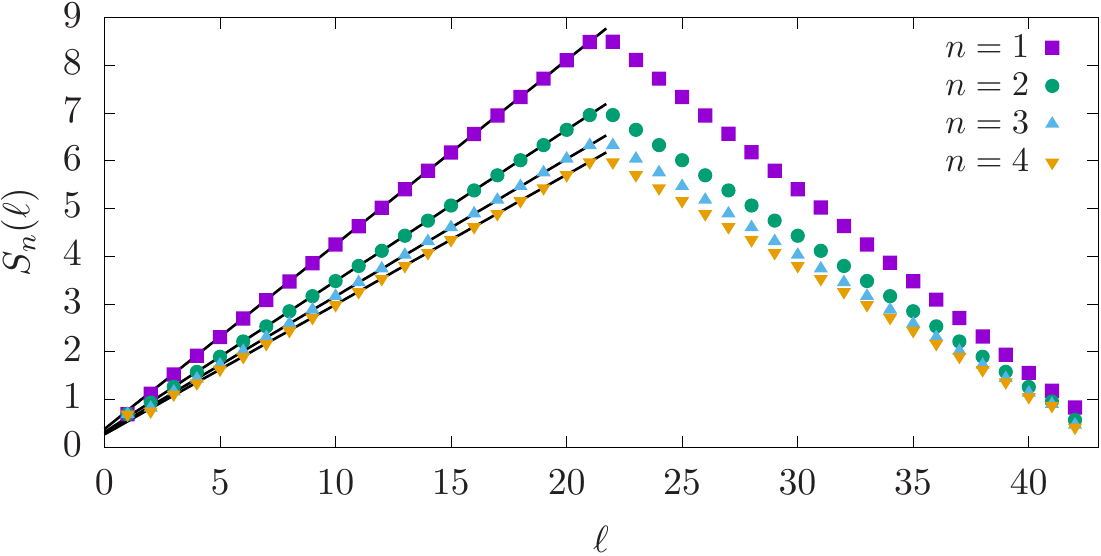} 
\caption{Different R\'enyi entropies for a scs system with $L=21$ and
 $h=9.2$ Black lines correspond to the theoretical expressions,
 Eq. \eqref{eq:renyis}.}
\label{fig:renyi}
\end{figure}

The R\'enyi entropies, defined as

\beq
S_m={1\over 1-m}\log\(\textrm{Tr} \rho^m\),
\label{eq:renyi}
\eeq
can also be computed:

\begin{eqnarray}
\nonumber
\label{eq:renyis}
S_2(l)&=&(l+1)\log \(24-16\sqrt{2}\),\\
S_3(l)&=&(l+1) 2 \log \frac{4}{3},\\\nonumber
S_4(l)&=&(l+1)\left(7\log{2}-\log \(17+8\sqrt{2}+4\sqrt{26+17\sqrt{2}}\)\right),
\end{eqnarray}
and these expressions can be seen to match perfectly the numerical
data for a scs rainbow chain with $h=9.2$ and $L=21$ in
Fig. \ref{fig:renyi}.

\section{XXZ Inhomogeneous model.}

\subsection{Hamiltonian}
\label{sub:subsection_name}

In this appendix we develop an alternative renormalization approach
based on the spin formalism which can be extended to the inhomogeneous
XXZ model, described by:

\begin{equation}
H=\sum_{n=1}^{2L} J_n \(S^+_nS^-_{n+1}+S^-_nS^+_{n+1} +
\frac{\Delta}{2}S_n^zS_{n+1}^z \) 
\equiv \sum_{n=1}^{2L}J_n(\mathbf{S_n}\cdot\mathbf{S_{n+1}})_\Delta
\equiv \sum_{n=1}^{2L}h_n,
\label{eq:ham_xxz}
\end{equation}
where $J_n$ follow the rule expressed in Eq. \eqref{eq:Jvalues}. On
the regime $h\gg 1$ it is natural to consider only the 3 spins which
are coupled with the strongest hopping amplitudes,
i.e. $n=L-1,L,L+1$. It can be checked that the GS of this Hamiltonian
lies on the sector of the total spin $S^{tot}_z=\frac{1}{2}$, so that
it is natural to renormalize the 3 spin block to an effective spin
$\frac{1}{2}$, $\mathbf{S_L^{(1)}}$.

\begin{eqnarray}
\ket{\tilde{+}}&=&\frac{1}{\mathcal{N}}(\ket{++-}-\lambda\ket{+-+}+\ket{-++}),\\
\ket{\tilde{-}}&=&\frac{1}{\mathcal{N}}(-\ket{--+}+\lambda\ket{-+-}-\ket{+--}),
\label{eq:states}
\end{eqnarray}
where

\begin{equation}
\mathcal{N}=\sqrt{\lambda^2+2}, \quad
\lambda=\frac{1}{2}(\Delta+\sqrt{\Delta^2+8}),
\end{equation}
and the GS energy is $E_0=-\frac{\lambda}{2}$. In order to determine
how the spins $\mathbf{S}_{L-1}$ and $\mathbf{S}_{L+1}$ got
renormalized we employ the Wigner-Eckhart theorem:

\begin{equation}
\mel{\tilde{m}}{{S^a_i}}{\tilde{n}}=\xi^a_i\mel{\tilde{m}}{S_L^{a(1)}}{\tilde{n}},
\end{equation}
which leads to:

\begin{eqnarray}
\xi^\pm_{L-1}=\xi^\pm_{L+1}=
\frac{2\lambda}{\mathcal{N}^2}, \qquad \xi^\pm_{L}=-\frac{2}{\mathcal{N}^2}, \\
\xi^z_{L-1}=\xi^z_{L+1}=
\frac{\lambda^2}{\mathcal{N}^2},\qquad \xi^z_{L}=\frac{2-\lambda^2}{\mathcal{N}^2}.
\label{eq:xis}
\end{eqnarray}

Note that $\sum_n\xi^z_n=1 \quad\forall\lambda$ and that
$\sum_n\xi^\pm_n=\frac{4\lambda-2}{\mathcal{N}^2}$, which is $1$ only
if the $SU(2)$ symmetry is present. This only holds when $\lambda=2$,
i.e. $\Delta=1$.

Hence we see that each step of the renormalization involves three
spins: $s_u$ and $s_d$ at the edges of the block and one central $s_c$
which is the outcome of each step except the first one, which is
physical too (see Fig. \ref{fig:illust} (b)).

\subsection{Fixed Points and RG flow} 
\label{sub:fixedpoints}

The next step of the renormalization procedure involves the spins
$S_{L\pm2}$. Using Eq. \eqref{eq:xis} we notice that the corresponding
terms of the Hamiltonian, Eq. \eqref{eq:ham_xxz}, can be written in
terms of the effective spin of the previous step:

\begin{equation}
 e^{-h}((\mathbf{S_{L-2}}\cdot\mathbf{S_{L-1}})_\Delta+
 (\mathbf{S_{L+1}}\cdot\mathbf{S_{L+2}})_\Delta) = 
 e^{-h}((\mathbf{S_{L-2}}\cdot\mathbf{S^{(1)}_{L}})_{\Delta'}+
 (\mathbf{S^{(1)}_{L}}\cdot\mathbf{S_{L+2}})_{\Delta'}),
\end{equation}
where

\begin{equation}
\Delta'=\frac{\Delta}{4}(\Delta+\sqrt{8+\Delta^2}).
\end{equation}

\begin{figure}[h!]
\includegraphics[width=80mm, angle = 0]{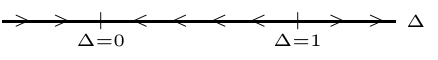}
\caption{RG flow  in terms of the anisotropy parameter $\Delta$.}
\label{fig:rgflow}
\end{figure}

Imposing $\Delta'=\Delta$, we determine the existence of two fixed
points: $\Delta_f=0$ (XX) and $\Delta_f=1$ (AFH). Futhermore,
iterating this equation while replacing
$\Delta\rightarrow\Delta\pm\epsilon$ with $\epsilon\ll1$ it is
straightforward to see that the former is stable
($|\Delta'|<|\Delta|,\Delta<1$) while the latter is unstable
($\Delta'\geq\Delta, \Delta\geq1$), as is depicted on
Fig. \ref{fig:rgflow}.

\subsection{MPS form}

As it has been described along the RG procedure, each step only
considers three spins: two physical ones, $s_u,s_d$ placed on the
edges of the block, and an effective spin, $s_c$ placed on the center,
which arises from the previous step. These three spins are
renormalized into a new effective spin $\frac{1}{2}$, $s'_c$. Hence,
Eq. \eqref{eq:states} can be written compactly in the form:

\begin{equation}
\label{eq:gs_mps}
\ket{s'_c}=\sum_{s_u,s_c,s_d}A^{s'_c}_{s_us_c s_d}\ket{s_us_c s_d},
\end{equation}
with

\begin{equation}
A^{s}_{\bar{s}ss}=A^{s}_{ss\bar{s}}=\frac{s}{\mathcal{N}},\qquad
A^{s}_{s\bar{s}s}=-\frac{s\lambda}{\mathcal{N}},
\qquad s=\pm, \bar{s}=-s.
\end{equation}

We can rewrite this elementary block of the MPS in another more
familiar form where central spins are now indices of the auxiliary
space.

\begin{equation}
A^{s'_c}_{s_us_c s_d}\rightarrow A^{s_u+s_d}_{s_cs'_c},
\end{equation}
so that

\begin{equation}
A^-=\frac{\lambda}{\mathcal{N}}\left(\begin{matrix}0 & 0 \\
1 & 0\end{matrix}\right),\qquad
A^+=\frac{\lambda}{\mathcal{N}}\left(\begin{matrix}0 & -1 \\
0 & 0\end{matrix}\right),\qquad
A^0=\frac{1}{\mathcal{N}}\left(\begin{matrix}1 & 0 \\
0 & -1\end{matrix}\right).
\end{equation}

If we particularize for the Heisenberg model, we recover (up to an
overall constant) the usual matrices that describe the MPS form of the
AKLT state \cite{Schollwoeck2011}. It is straighforward also to build
the basis used in \cite{P10,P12} to prove the degeneracy of the ES due
to the presence of the time reversal symmetry.

\end{document}